\documentclass[vecphys]{svmult}
\usepackage{ergebnisse} 
\usepackage{graphicx}
\begin{document}

\title*{Formation and Evolution of Supermassive Black Holes}
\author{Francoise Combes}
\institute{Observatoire de Paris, LERMA, 61 Av. de l'Observatoire,
F-75 014, Paris, France, 
\texttt{francoise.combes@obspm.fr}}

\maketitle
\begin{abstract} 
The correlation between the mass of supermassive black holes in galaxy 
nuclei and the mass of the galaxy spheroids or bulges (or more precisely
their central velocity dispersion), suggests a common formation scenario 
for galaxies and their central black holes. The growth of bulges and black 
holes can commonly proceed through external gas accretion or 
hierarchical mergers, and are both related to starbursts. Internal 
dynamical processes control and regulate the rate of mass accretion.  
Self-regulation and feedback are the key of the correlation. It is possible 
that the growth of one component, either BH or bulge, takes over, 
breaking the correlation, as in Narrow Line Seyfert 1 objects.
The formation of supermassive black holes can begin early in the universe,
from the collapse of Population III, and then through gas accretion.
The active black holes can then
play a significant role in the re-ionization of the universe. 
The nuclear activity is now frequently invoked as a feedback to star 
formation in galaxies, and even more spectacularly in cooling flows.
The growth of SMBH is certainly there self-regulated.
SMBHs perturb their local environment, and the mergers of
binary SMBHs help to heat and destroy central stellar cusps.
The interpretation of the X-ray background yields important constraints 
on the history of AGN activity and obscuration, and the census of AGN at low
and at high redshifts reveals the downsizing effect, already observed for
 star formation.  History appears quite different for bright QSO
and low-luminosity AGN: the first grow rapidly at high z, and their
number density decreases then sharply, while the density
of low-luminosity objects peaks more recently, and then 
decreases smoothly.  
\end{abstract}

\section{Introduction}
\begin{figure}[h]
   \centering
   \includegraphics[width=9cm]{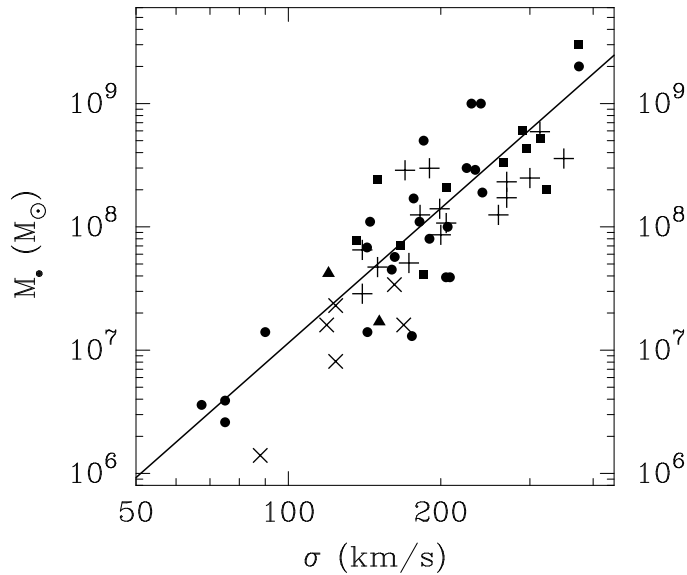}
      \caption{The relation between BH mass and
the velocity dispersion $\sigma$
inside the effective radius of the bulge. Filled circles indicate BH mass measurement
from stellar dynamics, squares from ionized gas, triangles on maser lines,
crosses are from reverberation mapping, and ``plus signs'' from ionization models
(from Kormendy \& Gebhardt 2001). The relation is close to a power-law of 
slope 4.}
       \label{fig1}
   \end{figure}

  It is now well established that all nearby galaxies possessing a 
spheroidal stellar component or bulge possess a central black hole (BH), 
with a BH mass proportional to the bulge mass, with a proportionality
factor which is now renormalised around 2 10$^{-3}$ (Magorrian
et al 1998, Gebhardt et al 2000, Merritt \& Ferrarese 2001, Shields 
et al 2003).
It soon appeared that the relation is more precise and with less scatter, 
between the BH mass and the central velocity dispersion (or dispersion 
inside the effective radius of the bulge $\sigma_e$), as shown in Figure \ref{fig1}.
The BH-mass grows then close to the 4th power of the central 
velocity dispersion.

The determination of this relation has been carried out by various methods:
\begin{itemize}

\item stellar proper motions for the Galactic center BH (Sch\"odel et al. 2003, Ghez et al. 2003),
\item stellar absorption lines, to obtain the stellar kinematics,
\item ionized gas emission lines (less reliable, since affected by outflows, inflows), and
also masing gas emission lines,
\item reverberation mapping, exploiting time delays between variations of AGN continuum,
and broad line emission, giving the size of the emitting gas region, combined with
the gas Doppler velocity to give the virial mass (Peterson \& Wandel, 2000)
\item ionization models: method based on the correlation between
quasar luminosity and the size of the Broad Line Region (BLR, Rokaki et al 1992).

\end{itemize}

The relation has recently been somewhat extended to lower masses,
in dwarf Seyfert 1 nuclei, 
which are more difficult to measure (Barth et al 2004, 2005).
Some progress has also been made in the search of intermediate mass black holes
(IMBH), for example in the globular clusters M15 in our Galaxy and G1 in M31:
in M15, the mass of the central object is lower than 10$^3$ M$_\odot$ and could be
stellar remnants (van der Marel 2003), while in G1, a BH of 2 10$^4$ M$_\odot$
is identified, and obeys the M$_{bh}-\sigma$ relation (Gebhardt et al 2002).
  A 1000 M$_\odot$  IMBH has also been estimated as member of a binary,
at the origin of the ULX source M82 X-1 in the starburst galaxy M82 (Portegies Zwart
et al 2004a).

The demography of SMBH, statistics on their activity frequency,
and their observed mass functions, constrain the possible AGN life-time
and growth rate.  It was already
suspected that AGN were active during a
short duty cycle of $\sim$ 4 10$^7$ yr,
and that many galaxies today should host a starving
black hole (Haehnelt \& Rees 1993). The observed M$_{bh}-\sigma$ relation now
strongly constrains the duty cycle time-scale. Also the cosmic background radiation
detected at many wavelengths constrains the formation
history. 
The volumic density of massive black holes today is derived,
from the observed density of bulges, and the 
proportionality factor M$_{bh}$ = 0.002 M$_{bulge}$.
And independently, the light that should have been radiated at
the formation of these BHs can be computed, redshifted and compared to the observed
cosmic background radiation: in the optical, we see only 10\% of the expected
flux, but 30\% in X-rays, and 80\% in the infra-red. 
The accretion radiation does not get out in optical light,
probably due to the extinction.

\section{BH growth}

Powerful QSOs are observed early in the universe, at z$>$6, 
with luminosities indicating very high BH masses, meaning
that masses as high as 10$^8$-10$^9$ M$_\odot$ can grow in less
than one Gyr.  However, the time-scale to grow a black hole
from a stellar mass of 10 M$_\odot$ to the Hills limit, M$_c$ 
(M$_c$ = 3 10$^8$ M$_\odot$), above which stars are
swallowed by the black hole without any gas radiation,
is of the order of 1.6 Gyr, if the gas accretion occurs
at the Eddington limit, and the efficiency is 
$\epsilon \sim 0.1-0.2$ (Hills 1975).
  The problem is therefore to accelerate the growth rate,
or begin from a higher initial mass.

\subsection{Quantifying the problem}

To have an order of magnitude,  and simple dimensional relations,
let us assume spherical accretion, 
from an accretion radius R$_{acc}$ = 0.3 M$_6$/v$_2^2$ pc,
where M$_6$ is the mass of the BH in 10$^6$  M$_\odot$, and
v$_2$ the velocity in 100km/s (corresponding to the effective stellar velocity
inside the galaxy nucleus, related to the bulge mass). The
canonical Bondi accretion rate is then:
$$dM/dt = 4 \pi R^2 v \rho = 10^{-4} M_\odot/yr M_6^2/v_2^3 \rho$$
where $\rho$ is the local density in M$_\odot$/pc$^3$.

Since dM/dt $\propto$ M$^2$, then the accretion time is $\propto$ 1/M,
t$_{acc}$ $\sim$ 10$^{10}$ yr/M$_6$ v$_2^3$/$\rho$;
for very low mass BH, this takes much larger than the Hubble time.
Therefore the formation of SMBH requires a large seed, mergers of BH, or very large
densities, like that characteristic of the Milky Way nucleus, 10$^7$ M$_\odot$/pc$^3$.

If these conditions are fulfilled, the growth of massive BH can then 
be accretion-dominated, i.e. t$_{growth}$ = t$_{acc}$. This phase could
correspond to moderate AGN, like Seyferts, and the luminosity is
increasing as  L $\propto$ dM/dt $\propto$  M$^2$. At some point,
the luminosity will reach the Eddington luminosity, since L$_{edd} \propto$ M.
The Eddington ratio increases as 
L/L$_{edd} \propto$ M, the BH growth slows down when approaching L$_{edd}$,
corresponding to a QSO phase. The time-scale of this powerful
AGN phase is
t$_{edd}$ = M/(dM/dt)$_{edd}$ = 4.5 10$^7$ yr (0.1/$\epsilon$)
(where $\epsilon$ is the usual radiation efficiency).
Equating t$_{acc}$ = t$_{edd}$, this occurs for 
M = 2 10$^8$ M$_\odot$ v$_2^3$/$\rho$ ($\epsilon$/0.1).
Wang et al. (2000) propose that tidal perturbations help to grow
a SMBH from a small seed, by boosting the accretion, and then
lead to the M$_{bh}-\sigma$ relation.

\subsection{Formation of the first massive black holes, in the early universe}

One solution to the growth problem could be that massive BH
form very early at high redshift, as the remnants of Pop III stars.
In the CDM scenario of hierarchical structure formation, it is 
generally thought that the first stars are expected to form in dark 
matter minihalos of mass 10$^6 M_\odot$, at redshifts around 20. Their 
virial temperature is too small for atomic hydrogen cooling to be efficient,
 but the molecular hydrogen cooling is fast enough (Tegmark et al 1997). 
Without metals and dust, the H$_2$ molecules form through H$^-$ with the 
electrons as catalysts. The minimum halo mass at a given redshift, in which 
the baryons are able to cool and form stars is obtained through the
 condition that the cooling time is smaller than the dynamical time, and is confirmed 
to be 10$^6 M_\odot$, 
at z $\approx$ 20-30 (Fuller \& Couchman 2000). Both 
semi-analytical estimations, and full 3D numerical simulations concord to
 find very massive first stars, with M* $> 100 M_\odot$ (e.g. Abel et al 
2002, Bromm \& Larson 2004).  Fragmentation is quite inefficient for these first condensations,
 due to the low metallicity and negligible radiative losses. The mass spectrum of these first
 stars is still not well known, but according to the 
cosmology, it is expected that the most massive structures are significantly clustered.

Above 260 M$_\odot$, the formed objects could collapse to a BH directly
(Bond et al 1984, Madau \& Rees 2001, Schneider et al. 2002). After the first subhaloes
have merged in larger entities, and formed dwarf galaxies, there could exist
10$^5$ M$_\odot$ IMBH in each center, formed by the merging of these seeds.

The total mass in these first black holes can be quite important. If every
halo corresponding to a 3 sigma peak (or higher) at z=24 forms a 260 M$_\odot$
BH, then the density per comoving volume is estimated at 
$\rho_{\bullet} = 2.9 10^5$ M$_\odot$/Mpc$^3$, already half of the present 
SMBH density (Islam et al 2004). 
It will then be sufficient to add some gas accretion to grow the BH along
their lives, and to ensure the merging of all seeds.
The problem at this stage is however the low efficiency of dynamical friction 
for objects that are still not massive enough. The consequence is that 
many BH will keep orbiting around subhaloes, instead of sinking to the main center.
Semi-analytical merger-tree calculations have been carried out, taking into account 
dynamical friction, tidal disruption and encounters with the galactic disk, to determine 
the abundance and distribution of MBHs in present-day haloes of various masses (Islam 
et al 2004): the result is that it is difficult to 
reproduce the observed mass distribution of SMBH with only merging of the initial seeds, 
and that further gas accretion is required. Also the formation of
binaries at the center of structures require gas accretion in order
for the binary BH to merge before a triple is formed and some BH are lost in 
intergalactic space.

\begin{figure}[h]
   \centering
   \includegraphics[width=9cm]{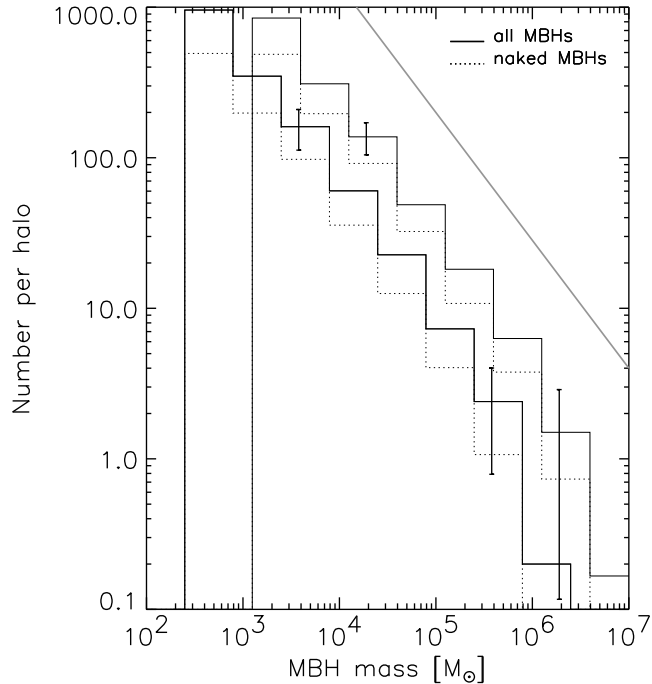}
      \caption{ The mass distribution of massive BH (naked, i.e.
without any satellite halo, or not) predicted in 
a final halo of 1.2 10$^{12} M_\odot$, for two hypothesis on the seed mass
(260 and 1300 M$_\odot$, for the upper and lower curves respectively),
from Islam et al. (2003). }
       \label{islam}
   \end{figure}

The consequence of low merging efficiency of the seed BH is the predicted large 
abundance of these intermediate mass BH in a given galactic halo (cf Figure 
\ref{islam}): typically a thousand or more should exist in the Milky Way.
Coming from rare high density peaks, they are expected to cluster
in the bulges and spheroids; when they accrete gas, they could account for
ultra-luminous X-ray sources (ULX) which are offset from the galaxy centers.
In particular, masses typical of large IMBH, i.e. 10$^5 M_\odot$, should reach the
number of $\sim$ 10 in the Galaxy. Also, it is found that hierarchical merging
can only be responsible of 10\% of the total mass of present SMBH,
and that gas accretion should be responsible for the rest. 
Taking into account the progressive gas accretion along the BH growth
leads to a present SMBH density comparable to what is observed
(Volonteri et al 2003a), and also to a large number of wandering IMBHs. 

Numerical simulations show that the M$_{bh}-\sigma$ relation can 
indeed be conserved through several successive mergers, provided that
gas dissipation and star formation is included at each merger
(Kazantzidis et al 2005); collisionless mergers could cause
some scatter in the relation.

\subsection{Mini-quasars and reionization}

An intense UV background is expected from the first stars.
These stars are so massive that they create an HII region around them,
able to blow away all the gas from the mini-halo where they 
are seated (Whalen et al 2004). The UV photons in the Lyman-Werner bands
are able to photo-dissociate the fragile H$_2$ molecules
in the neighbourhood, preventing the gas to cool down.
Star formation will be inhibited in a large
region, until accumulation of gas creates dense
regions able to be shielded. 

At the death of the first stars, the massive black holes
created from merging of the seeds,
could accrete gas and become mini-quasars, able to produce
a harder radiation background, including hard and soft X-rays.
It has been argued that these X-rays could have a positive
feedback on the formation of the H$_2$ molecules in producing
electrons, and compensate for the negative feedback of the
UV background (Haiman et al 2000).
However, 3D detailed simulations, find that the positive
feedback is barely sufficient (Machacek et al 2003).

A fundamental question is to know precisely at which
epoch the inter-galactic hydrogen has been completely
re-ionized, ending the dark age, and whether this has been
done essentially through stellar radiation of from mini-quasars.
The discovery of the Gunn-Peterson trough in some z$>$6
quasars of the Sloan Survey (Becker et al 2001) suggests that
reionization is occurring near z=6, while the WMAP result
of a high electron scattering optical depth implies that ionizing
sources were present up to z=15, suggesting a long reionization period,
may be in two steps (very massive stars at z=15, and after a feedback
epoch, much less massive stars at z=6). The possibility of mini-quasars
as the source of reionization has been studied by Dijkstra et al (2004), in view
of the X-ray background constraints. The hard X-ray photons produced by
the miniquasars would be observed today as a soft X-ray background.
If the quasars were only responsible for the reionization, than
They will overproduce X-rays, and be incompatible with the observed 
0.5-2 keV background. The miniquasars could only be responsible to about 50\%
 of the IGM reionization.

\subsection{The case of IMBH }

\begin{figure}[h]
   \centering
\includegraphics[width=10cm]{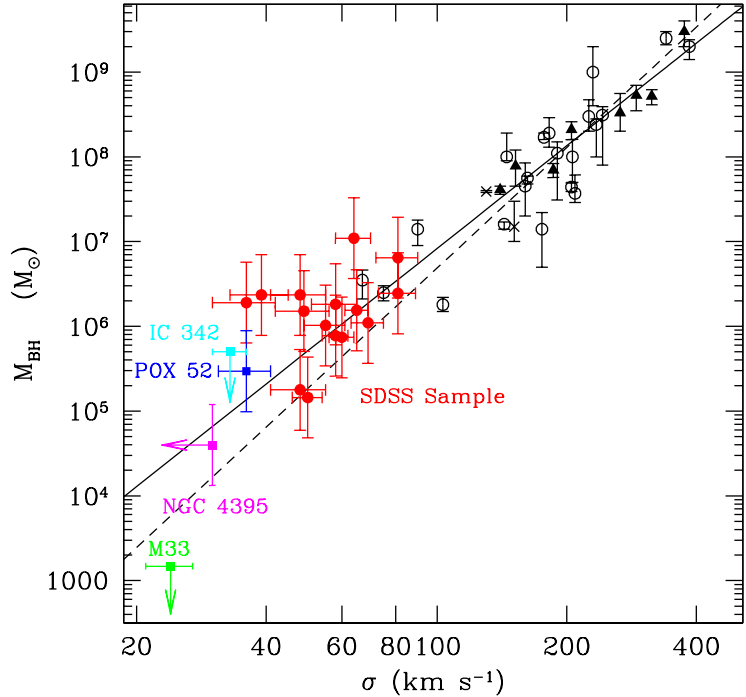}
\caption{ Extension of the M$_{bh}-\sigma$ relation to IMBH
(from Barth et al. 2005).
  Points in the upper right are black holes with dynamical mass
  measurements compiled by Tremaine et al (2002).  Open circles represent
  stellar-dynamical measurements, filled triangles are gas-dynamical
  detections, and crosses are from H$_2$O maser observations. In the 
  middle of the diagram,  
  filled circles are the SDSS Seyfert 1 galaxies  studied by Barth et al
(2005).  The low mass points are indexed by their galaxy name.
 Note the upper limit for any black hole mass in M33 by 
Gebhardt et al (2001).}
       \label{barth}
   \end{figure}

Does the M$_{bh}-\sigma$ relation extrapolates to low masses?
At least below 10$^6 M_\odot$, the extrapolation appears straight
forward (Barth et al 2005), however, it is difficult to bridge the
gap towards the low end
of  intermediate mass black holes (of 10$^3 M_\odot$);
 their observation
is very difficult, both by the kinematics, since their gravitational influence
is small, and from their possible AGN activity, since the expected
luminosity is weak. According to the extrapolation of the M$_{bh}-\sigma$ relation, these IMBH
should be searched as AGN in dwarf galaxies: among the good
candidates are NGC 4395 (Filippenko \& Ho 2003), where the BH mass
is likely to be 10$^4$-10$^5$ M$_\odot$ (radiating much below
the Eddington limit), or Pox 52,
with $\sim  2 10^5 M_\odot$ (Barth et al 2004).
The problem of this search is that dwarf galaxies
frequently host nuclear star clusters of $\sim$ 10$^6$ M$_\odot$,
hiding the weak AGN. They are best observed in the
Local Group; a famous example, M33, does not host any BH
more massive than  10$^3$ M$_\odot$, which is already 10
times below the value expected from the M$_{bh}-\sigma$ relation
(cf figure \ref{barth}).

One of the main features of lower mass IMBH is that they are no longer
unique objects at the very center of the galaxies, but could 
be found in large numbers, spread out in galaxy haloes.
The dynamical friction is no longer efficient enough, and 
their formation mechanism, through mergers of lower
mass black holes through binary coalescence could
provide them randomly large velocities (Ebisuzaki et al 2001).

X-ray observations have revealed in nearby galaxies a class of extranuclear  point sources with 
X-ray luminosities of 10$^{39}$-10$^{41}$  ergs/s, exceeding the Eddington luminosity for stellar mass 
objects. These ultraluminous X-ray sources (ULXs) may be powered by 
intermediate-mass black holes of a few thousand M$_\odot$  or stellar mass black holes 
with special radiation processes.

Liu \& Bregman (2005) find a strong association between ULXs and star formation sites,
ULXs are preferentially observed in late-type galaxies, in spiral arms, and in some cases
associated with supernovae. However a few ULXs are observed in old globular clusters,
and Colbert et al (2004) find them associated with population II stars, in particular in elliptical 
galaxies.

These ULX sources could be due to intermediate black holes resulting from
the mergers of massive BH seeds, formed out of Population III objects in the early universe
(Volonteri \& Perna 2005). The large number of BH formed can merge in galaxies,
through binary coalescence, but the possibility of a triple association, followed by
the ejection of one of the BH, and recoil of the binary, leads to the prediciton of many
IMBHs wandering throuh the haloes. To be observed radiating at the high end of X-ray luminosity,
these sources must be associated to baryons, and the most probable locations
are in the disk of late-type galaxies (Volonteri \& Perna 2005). 

There are however observational problems in the interpretation of ULXs
in terms of IMBH (Makishima et al 2000). In most of
them, the inner disk temperature is observed around 2kev, too high 
to be compatible with the high black hole mass, as required with the IMBH
hypothesis, radiating at Eddington luminosity.
They might be a mixed-bag class
of objects, some could result from beamed emission during a short phase of
common X-ray binaries, and they could be related to micro-quasars.
This would explain their relation to star formation region. 
(e.g. King et al 2001).    Some could represent the intermediate mass
black hole, expected in the continuity of SMBHs. An interesting case
is the ULX source right at the nucleus of M33 which nature is still debated
(Dubus \& Rutledge 2002).  The central source corresponds 
to radiosources expected for micro-quasars (Trejo et al 2004).

Evidence for an IMBH could come from the Milky Way nucleus:
Hansen \& Milosavljevic (2003) propose its existence to explain
the observation of 
bright stars orbiting within 0.1pc, which are are young
massive main-sequence stars, in spite of an environment hostile
to star-formation. Aternative solutions could be
star mergers, or exotic objects (Ghez et al. 2003).
In the IMBH scenario, stars were formed in a star cluster 
outside the central pc, and then dragged in by
a BH of 10$^3$-10$^4$ M$_\odot$. The decay time-scale by
dynamical friction for normal
stars is too large (much longer than the massive stars life-time),
but for the IMBH, this time-scale is 1-10 Myr.
Stars may be dragged inwards even after the star cluster 
has been disrupted.

Such a system SMBH-IMBH and a gas disk may reveal interesting
dynamics; it is  similar to a protosolar system,
with the Sun-Jupiter couple, resonant effects like
planetary migration are expected (Gould \& Rix 2000).

\subsection{NLS1 and black hole growth}

The M$_{bh}-\sigma$ relation has been established locally, and it is not
yet known whether the relation was already there in the primordial structures, and 
then was maintained during the evolution by a feedback process, or was obtained 
progressively, without maintaining a permanent relation.  

There might be phases in the life of a galaxy, where the star formation has some advance 
with respect to the black hole growth, according to the various feedback and regulating
 mechanisms, and we should be able to recognize a sub-class of AGN where the BH-mass 
is somewhat below the standard relation. This has been proposed by Mathur (2000) for
 Narrow Line Seyfert 1 galaxies (NLS1). Grupe \& Mathur (2004) investigated the BBR
 relation for a sample of broad-line Seyfert 1 galaxies (BLS1s) and narrow-line Seyfert
 1 galaxies (NLS1s), and confirm that NLS1s lie below the BBR relation of BLS1s. 
As a consequence, black holes grow by accretion in well-formed bulges, possibly after 
a major merger. As they grow, they get closer to the BBR relation for normal galaxies
(Mathur \& Grupe 2005).
 The accretion is highest in the beginning and then decreases with time. There is no
 AGN feedback for the control of bulge growth there. Kawaguchi et al (2004) estimate 
that the NLS1 phase is characterized by very efficient accretion, at a super-Eddington
 rate; given the high frequency of these objects (10 of all AGN), and the
 average duty cycle for an average AGN phase ($10^7-10^8$ yrs), the essential of the
 BH growth is occurring during this phase: the BH grows by up to 1-2 orders of magnitude,
 while in the BLR phase (the most frequent and common phase) at sub-Eddington rate,
 the BH will only multiply its mass by about 2. 

\begin{figure}[h]
   \centering
\includegraphics[width=10cm]{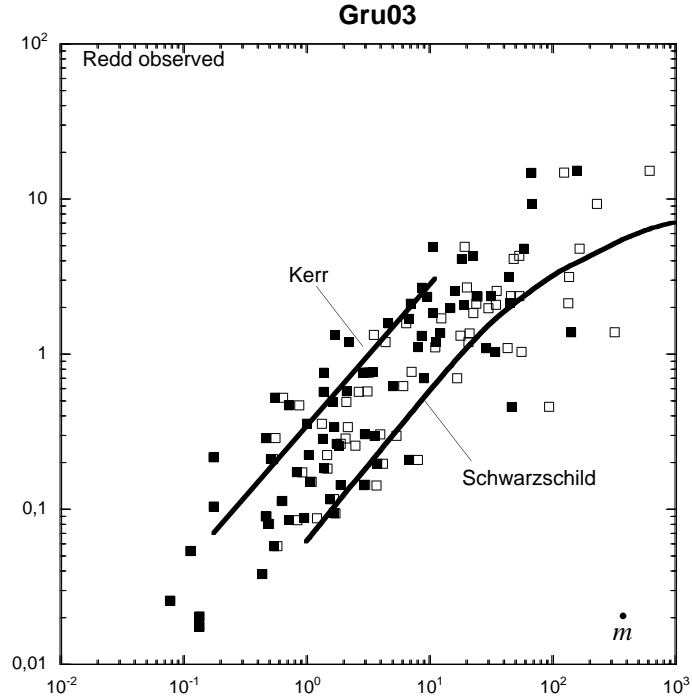}
\caption{The Eddington luminosity ratio $R{\rm edd}$ as a function of $\dot{m}$, 
the accretion rate in Eddington units,  $\dot{m} = \dot{M} c^2 /L_{\rm edd}$,
for the NLS1 sample of Grupe \& Mathur (2004). The luminosities are
computed with the standard disc (open squares) and the self-gravitating
disc (filled squares).  The two curves correspond
to the slim disc model,  and respectively a Schwarzshild
and a Kerr BH (from Collin \& Kawaguchi 2004).
}
       \label{collin}
   \end{figure}

When the accretion rate is much larger than Eddington, the accretion is occuring not
through a "thin" but a "slim" disk, with a cooling time larger than the viscous
time, so that energy is advected towards the BH before being radiated.
The luminosity can then saturate, and never be larger
than a few Eddington luminosity. 
According to the type of BH, the accretion rate can range from 60 (Schwarzschild BH)
to 300 Eddington accretion rate (Kerr BH). Whatever these accretion rates,
and whatever the mass of the BH, the luminosity is always no more than 10 Eddington
luminosities, as shown in Figure \ref{collin},
and the accretion rate a few M$\odot$/yr (Collin \& Kawaguchi 2004).
This is a strong indication of a mass-limited supply, with an external
mechanism to regulate the accretion.

\subsection{Down sizing, and life-time of activity}

It is now well known that the physics of baryons, both the
star formation, and gas accretion by black holes, act to
compensate the hierarchical formation of dark matter haloes,
which grow larger and larger with time: the most massive star-forming galaxies
and the most massive SMBH are forming at high redshift, early in the 
universe, while only smaller masses are assembling now (Cowie et al 1996). 
Semi-analytic follow up of these processes have succeeded to reproduce
the down sizing, and taking into account the constraints of luminosity 
functions of galaxies and AGN at all redshifts can teach us more on the
formation of the objects, and for instance on the duty-cycles or life-time
of activity. 

\begin{figure}[h]
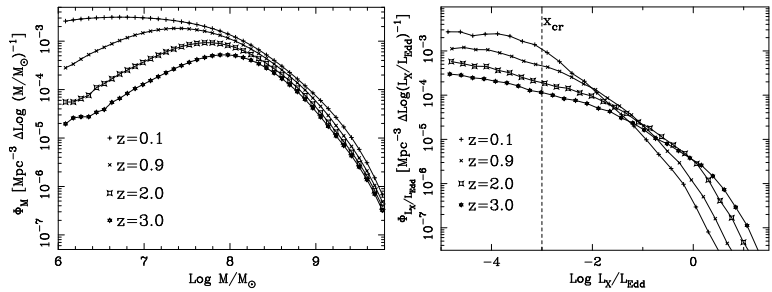

   \centering
\begin{tabular}{cc}
\includegraphics[angle=270,width=5cm]{smbh_fig5a.ps}&
\includegraphics[angle=270,width=5cm]{smbh_fig5b.ps}\\
\end{tabular}
\caption{ Illustration of the down sizing effect in growing BH:
{\bf Left}: Redshift evolution of the SMBH mass function, from z=3 (bottom) to 0.1 (top)
{\bf Right}:  corresponding evolution of the X-ray luminosity
(normalised to Eddington rate), from Merloni (2004).}
       \label{merloni}
   \end{figure}

Using the observed correlations between X-ray and radio 
luminosities of quasars and their black hole mass, Merloni (2004) has
computed the past history of SMBH, assuming their growth is only due to gas
accretion. The accretion rate, and radiative regime (Eddington or not)
is not fixed, but derived by the model. The results show a clear 
anti-hierarchical growth of the black holes, as shown in Figure \ref{merloni}.
 The most massive SMBH are in place at high redshift, while at low redshift
only smaller mass black holes are accreting, so that the average 
BH mass of observable AGN is increasing with redshift.
The life time of activity is also varying with redshift, being shorter
at early times. The mean life time is defined by the average over the activity of the
 time, weighted by the accretion rate. It is not imposed to be the 
doubling time of the mass at the Eddington rate, i.e. the Salpeter time
$ t_s = \epsilon M c^2 /L_{Edd}$ = $\frac{\epsilon}{0.1} 4.5 10^7$ yrs. The life time
 ranges from $10^7$ yrs to assemble $10^9 M_\odot$ at z=3, up to 
$10^8$ yrs to assemble $10^7 M_\odot$ at z=0 (Merloni 2004).

\subsection{Quasar life times}

The quasar life time can be estimated from the observations of
the AGN demography,
through the statistical argument that the fraction of active nuclei
among the whole number of SMBH present in every early-type galaxy
is a measure of the time spent in the active phase.
The observation of the proximity effect (presence of an ionized region around
the quasar host) puts already a lower limit to the life time of
10$^4$ yrs. The life time is defined as the total active time
of a quasar, i.e. if the same quasar experiences episodic activity, the
life time is the sum of the duration of the active phases.
The minimum life-time of 10$^4$ yrs is for one episode only.
The various methods to estimate the quasar life time have been 
reviewed by Martini (2004), and results in values between 10$^7$ and
10$^8$ yes, quite close to the Salpeter time, or mass doubling time,
assuming Eddington luminosity, and with efficiency $\epsilon$ = 0.1. 
One of the longest duration of an episode may be derived by the observed
length of radio jets, about $t_Q \sim 10^8$ yrs. 
Through the measurement of the AGN-galaxy cross-correlation length,
Adelberger \& Steidel (2005) conclude that high and low luminosity
AGN are both found in haloes of similar masses, and therefore the
higher observed frequency of faint AGN must imply that their duty cycle
is much longer than for bright AGN, of a few Gyr. 

 All these estimations are compatible to the hypothesis that the 
active phases of bright AGN
are triggered by a major merger between two gas rich galaxies, that
removes angular momentum and drives the gas towards the center
(Barnes \& Hernquist 1992). This hypothesis is supported by 
the frequent association between quasars and interactions (e.g. Hutchings
\& Neff 1992).
However, it is possible that the observed quasar life-time is 
biased in the observations, if the active phase, where the BH grows
and radiation is emitted, is partially obscured by dust, as expected
when a lot of gas is driven towards the galaxy centers, in the beginning
of the activity. Hopkins et al (2005) have estimated the importance
of this obscuration phase in numerical simulations, and find that
the quasar life time is then reduced from an intrinsic value of 100 Myr,
to an observable value of 10-20 Myr.

\section{Interpretation of the M$_{bh}-\sigma$ relation}

Several models have been proposed to account
for the relation, all involving a simultaneous formation
of bulges and SMBH, and constraining the
feedback processes.

\subsection{Radiative feedback}

 Although the radiative feedback is not the most efficient, it can
play an important role at the end of the feeding of a giant black hole in 
an elliptical galaxy, which by definition does not possess much gas. Sazonov
 et al (2004) have computed the equilibrium temperature $T_{eq}$ 
of the gas around
a quasar, heated by Compton scattering and photoionization, and cooled
 by continuum and line emission. When $T_{eq}$, which is proportional 
to $L/(nr^2)$ becomes larger to the virial temperature of the 
galaxy, proportional to the velocity dispersion $\sigma^2$, the gas is 
expelled, and the fueling is stopped. This occurs when the density $n$ 
becomes lower than a critical density, $n_{crit} \propto L / (r^2 \sigma^2)$.
Assuming that the gas distribution follows the stellar distribution, which is isothermal,
 with an $r^{-2}$ radial profile, then the equilibrium temperature is constant with radius,
 and inversely proportional to $\sigma^2$. At the critical regime, when $T_{eq} = T_{vir}$, 
the maximum BH-mass is then proportional to $\sigma^4$, and its growth is stopped. The
 radiative feedback then could explain the M$_{bh}-\sigma$ relation, for massive ellipticals,
 with very low gas mass content (Sazonov et al 2005).

\subsection{Feedback due to QSO outflows}

QSO and stars main cosmic formation epoch coincide
(e.g. Shaver et al. 1996). Their common formation
could be regulated by each other, and the QSO outflows
prevent star formation (Silk \& Rees 1998).
The condition for the wind to be powerful enough to
give escape velocity to the gas constrains
the BH mass to M$_{bh}$ $\propto$ $\sigma^5$,
which from the Faber-Jackson relation, gives  M$_{bh}$
 $\propto$  M$_{bulge}$.
But the phenomenon is assumed spherical, 
in reality jets are collimated, the gas is clumpy, 
and compressed to form stars.

\begin{figure}[h]
   \centering
   \includegraphics[width=9cm]{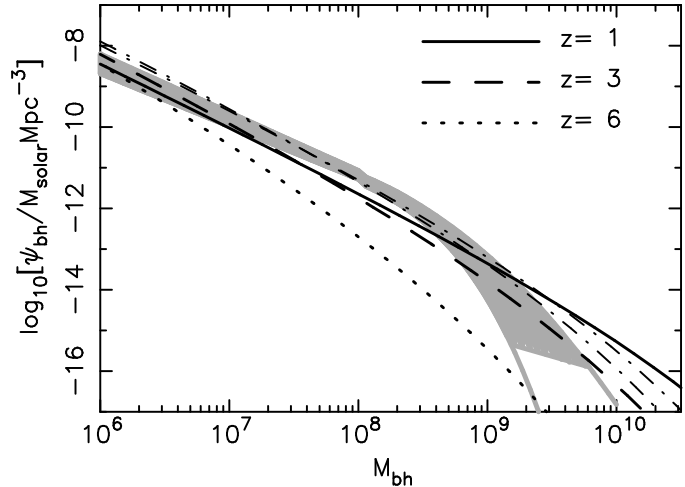}
      \caption{The mass function of SMBHs. The
grey region shows the mass function estimated from the velocity
function of Sheth et al.~(2003) and the M$_{bh}-\sigma$ relation
of Merritt \& Ferrarese~(2001). The three lines, solid, dash, dots
represent the computations at $z=1$, 3 and 6 from the Press-Schechter halo mass
function by Wyithe \& Loeb (2003). The dot-dashed lines show 
the mass function at
$z\sim2.35$ and $z\sim3$ implied by the observed density of quasars
and a quasar lifetime $t_{\rm dyn}$ (Wyithe \& Loeb 2003).}
       \label{loeb}
   \end{figure}

The mass of the SMBH in a Galaxy is quite negligible, 
lower than $10^{-3}$ in general, and the distance
to which the gravitational action is significant is 
quite small, less than 100pc even for the more massive BH. 
But the energy that the SMBH can radiate or expel as
jets, winds or outflows, is relatively large, in comparison
to the binding energy of the gas component in the Galaxy, and
therefore, the energy output of the AGN could have some
significant feedback action, in stopping the gas inflow
favoring the accretion (Wyithe \& Loeb 2002).
Let us note that the binding energy of the gas, rotating at 
circular velocity $v_c$ in the galaxy, is of the order of
$(v_c/c)^2 \sim 10^{-6}$ of its rest mass, while the energy output of the
AGN could be much larger, around 10\% of the rest mass energy,
so that the accretion of only a fraction  $10^{-5}$ of the gas
mass will be sufficient to release its binding energy.

This could explain the relation between the mass of the black
hole and the central velocity dispersion, as a self-regulation
mechanism. If the binding energy of the system of mass $M$ is of the order
$M \sigma^2$, and its dynamical time $r/\sigma$, the typical 
energy per dynamical time is $M \sigma^3/r$. Eliminating the mass
through the virial relation $M \sim \sigma^2 r/G$, the typical
energy rate or luminosity is $\sigma^5/G$. This can be considered as
the maximum luminosity of the black hole before unbinding the system
hosting it. Equating this to the Eddington luminosity, relates the
mass of the black hole to $\sigma^5$ with an order of magnitude
quite comparable to the M$_{bh}-\sigma$ relation observed
(Silk \& Rees 1998, Ciotti \& Ostriker 2001). The proportionality factor 
takes into account the low coupling of the energy of the quasar
(wind, outflows) to the galaxy gas. About 5-10\% of the energy of
the quasar must be absorbed by the galaxy to explain the self-
regulation. Also the self-regulation might account for the maximum
mass observed for SMBHs, which are never more massive than a few 10$^9
M_\odot$.

The principle of the self-regulation is welcome to account for the 
very short duty cycle of nuclear activity in galaxies. The 
statistics of the number of AGN with respect to the quiescent 
SMBH in all galaxies leads to a duty cycle as short as 10$^7$
to 10$^8$ yrs, according to the strength of the AGN. The duty cycle is of
the same order as the dynamical time of the gas feeding the AGN.

\subsection{Models based on self-regulation growth} 

The detailed computation has been done by several groups,
with different assumptions. The main lines are that the
BH grows as long as the energy released in the galaxy is
lower than the binding energy. If the heated gas can cool
with a sufficiently short time-scale, more energy is required for
the feedback, by a factor up to $c/\sigma$, and the resulting
relation is then M$_{bh} \propto \sigma^4$.
The mass function of quasars is obtained, assuming that the
BH grows in galaxy mergers (Kauffmann \& Haehnelt 2000), both
by the merging of the BH, and also by gas accretion infalling
during the interaction. The quasars are assumed to radiate at Eddington 
luminosity during their duty cycle, which is comparable
to the dynamical time of the feeding system. The peak in the
quasar luminosity function at $z \sim 2$ is obtained through the
merging history, since it coincides to the peak of the formation
of massive ellipticals, while the galaxy clusters are forming.
The maximum BH masses correspond to the maximum galaxy masses, 
obtained at these epochs. The models then should cut off the
gas infall in haloes with velocity larger than 500 km/s typically
(cf Figure \ref{loeb}).
These systems correspond to small clusters of galaxies, where the
hot gas cannot cool to fuel a central SMBH. The duty cycle of quasars
of 3 10$^7$ yrs corresponds also to the peak of quasar luminosity at
$z \sim 2$, but this time scale must be shorter at high redshifts.

The main conclusions of these models is that 80-90\% of the SMBH mass
has been already accreted at $z \sim 1.5$ (Wyithe \& Loeb 2003).
The total light in galaxies can be also modelled according to the
same ideas, assuming that star formation is regulated by feedback,
and ceases when an energy comparable to the binding energy is released
(Dekel \& Woo 2003). However, the dependence in redshift of the efficiencies
to accrete mass for black holes and star formation is not the same, and
therefore the M$_{bh}-\sigma$ relation should be z-dependent. Only the
M$_{bh} -\sigma$ relation should be constant. 
Indeed for a given dark halo mass, the dependence of
the stellar mass is in $(1+z)$ (constant characteristic
feedback time), while that of M$_{bh}$ is in $(1+z)^{5/2}$,
and their ratio is $\sim (1+z)^{3/2}$. The BH mass is larger 
with respect to the 
stellar mass at high redshift, with the same M$_{bh} -\sigma$ relation, 
since stellar systems are more centrally condensed at high $z$.

\begin{figure}[h]
   \centering
   \includegraphics[width=9cm]{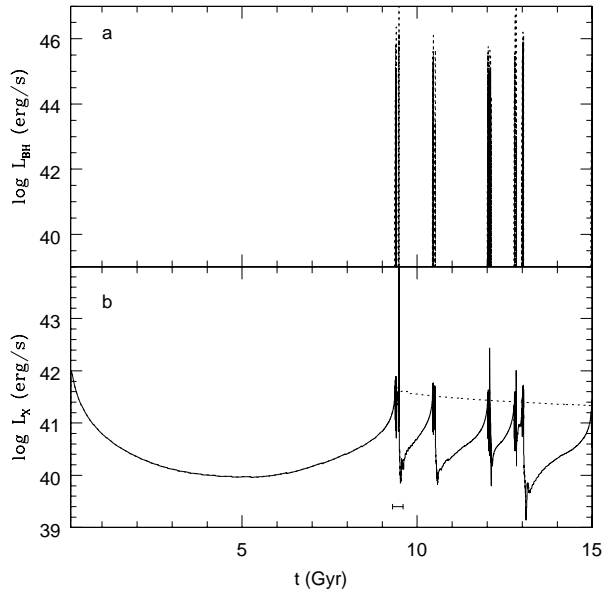}
      \caption{ Self-regulation due to the quasar in an elliptical galaxy:
      Top (a): the time evolution of $L_{bh}$ (bolometric) emitted by the quasar
      at the galaxy centre. Bottom (b): the time evolution of $L_x$ for
      the model with $\epsilon=0.1$ (solid line), and that of the same model
      with $\epsilon=0$ (cooling flow - dashed line), emitted by the whole galaxy,
     up to the  truncation radius, from Ciotti \& Ostriker (1997).}
       \label{ciotti}
   \end{figure}

The influence of AGN feedback due to energetic winds can be studied
through numerical simulations, adopting simple recipes for the accretion
of gas by the growing black holes, and energy release through 
winds in the interstellar medium. Di Matteo et al (2005) have
then compared major mergers between two spiral galaxies with and
without the presence of BH, and shown the dramatic difference
between the star formation rate, and the presence of gas in 
the remnant.  A series of such simulations, where the star formation 
rate and BH accretion rate are self-regulated, can yield at
the end the M$_{bh}-\sigma$ relation in the present remnants.

\subsection{AGN feedback and cooling flows}

The self-regulation between accretion and feedback appears to be at
work in elliptical galaxies, where the cooling of the gas in only intermittent,
and at larger scale in galaxy clusters, where huge cooling flows are
impeded through re-heating by the central AGN. Ciotti et al (1991) and 
Binney \& Tabor (1995) developed the regulating mechanism, based on the two 
opposed sources: mass loss from evolving stars fuels the galaxy in gas, and 
the heating by Type Ia supernovae keeps it far from the cold phase, but 
with a faster declining efficiency. Since the heating by supernovae cannot 
compensate for the mass
drop out, there must occur a cooling catastrophe, fueling the central black
hole now known to be present in every elliptical galaxy. The energy 
release during the short active phase reheats efficiently the gas, which 
is then the source of X-ray radiation. The intermittent AGN phases are
schematically shown in Figure  \ref{ciotti}, revealing relaxation oscillations.

\begin{figure}[h]
\centering
\includegraphics[width=10.5cm]{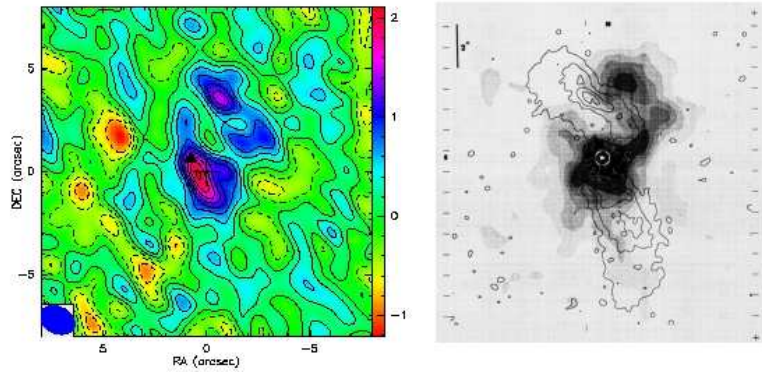}
\caption{ Cold gas associated to the Abell 1795 cooling flow:
{\bf Left} CO(2-1) map obtained with the IRAM interferometer,
from Salom\'e \& Combes (2004). The AGN is indicated by the white ring.
{\bf Right} H$\alpha$ +[NII] line emission (grey scale), with
6cm contours from van Breugel et al. (1984).}
\label{cool}
\end{figure}

At larger scales, it has become evident in recent years, thanks to the
progress of X-ray observations by Chandra and XMM-Newton, that cooling flows in 
galaxy clusters are completely different from the stationnary, symmetrical and
abundant phenomenon expected by simple theoretical ideas.  
The X-ray observations have constrained the amount of cool gas observed,
and the cooling rates have been reduced by at 
least one order of magnitude;
the old view of quiet and regular, quasi-spherical cooling
has given place to partial and intermittent cooling,
perturbed by re-heating and feedback processes due to the central AGN.
The compensation of cooling and heating could even be
used to measure the power of the AGN
(Churazov et al 2002).
A spectacular illustration of this perturbed
cooling is the Chandra image of the cooling flow in Perseus,
with bubbles, shocks, gas streaming up and down from the center,
and ripples looking like emitted sound waves (Fabian et al. 2003). In
the same time, cold gas in the form of CO molecules were observed in
dozens of cooling flows (Edge 2001, Salom\'e \& Combes 2003), and the
amount of cold gas corresponds to the order of magnitude expected
by the revised cooling rates. High spatial resolution observations show
that the cold gas is associated to the dense X-ray gas, compressed by
the AGN lobes, and is present around the cavity created by the lobes
(cf Figure \ref{cool}). In these dense regions, star formation occurs,
and HII regions are observed. 

All these new observations concord to draw a picture where the cooling
flows are intermittent, and the AGN feedback is self-regulating both the
growth of the central black hole mass, but also the amounts of
star formation in the central galaxy.

\subsection{Hierarchical models of galaxy formation}

The mass assembly of supermassive black holes in galaxies, simultaneous
to the build-up of their stars can well be integrated in the 
CDM scenario of hierarchical formation, and the
M$_{bh}-\sigma$ relation follows (Haehnelt \& Kauffmann 2000).
  The main hypotheses of the model are that black holes
are essentially assembled in galaxy major mergers, which 
simultaneously form starbursts, elliptical galaxies,
and fuel a QSO phase. An additional assumption is that the
fraction of gas transformed into stars per dynamical 
time increases along the Hubble time, while  the available
gas fraction in galaxies decreases, as does the rate of gas
accretion by galaxies.  The required gas fraction accreted 
by the black hole grows with the mass
of the halo, and the accretion time  scales with the dynamical time.

In this model, the scatter  in the M$_{bh}-\sigma$ relation is due to:
\begin{itemize}
\item  M$_{gas}$ of the bulge progenitor depends on $\sigma$, but not on
the formation epoch of the bulge, while M$_*$ depends on both; 

\item  mergers move the galaxies on the M$_{bh}- \sigma$  relation,
even at the end, when there is only BH mergers,
and not enough gas left to grow the black hole.
\end{itemize}

The gas fraction in galaxies falls from 75\% at z=3 to 10\%  at z=0.
The gas fraction in major mergers is higher in 
fainter spheroids that form at high z, which are more concentrated.
Elliptical/spheroids forming recently have smaller BH.

Typically a seed BH of 10$^6$ M$_\odot$ forms at 5 $<$ z $<$ 10 
and then gas is accreted. For a typical SMBH,
about 30 black holes are merged. Today black holes in big ellipticals accrete
only by merging with small galaxies, but in the past gas accretion
was dominant.

Both the number density of quasars as a function of redshift,
and the evolution of gas abundance are found compatible with observations.
The required duration of a QSO phase is 10$^7$ yrs
(Haehnelt \& Kauffmann 2000).

\subsection{Feedback through bars in spiral galaxies}

If quasars, which are the high luminosity end of the AGN population, 
are clearly associated with interactions and mergers
(Hutchings \& Morris 1995), it is difficult 
to trace evidence of dynamical triggering mechanisms for milder AGN, 
like Seyfert or LINERS in spiral galaxies (e.g. Combes 2003).

The accretion rates required are of course very different, of the order
of a few 10 M$_\odot$/yr for quasars, and more than two orders of 
magnitude less for nearby Seyferts, so that the dynamical processes are
much less violent, for Low Luminosity AGN (LLAGN).
However, most SMBH in galaxies today have been built by gas accretion, 
since the successive mergers of BH from the primordial ones are far
insufficient (Islam et al 2004), so it is of prime importance to understand
the dynamical processes responsible for gas accretion in the nearby
LLAGN, that can be studied in details.

Non-axisymmetries, and essentially bars, are the main providers of
gravity torques, that will make the gas lose its angular momentum, and
infall towards the center. This is the main mechanism both for isolated 
galaxies with spontaneous bar instabilities, and also during galaxy 
interactions, that favor bar instability: bars are then the way to propagate
tidal interactions in the inner parts of galaxies (e.g. Barnes \& Hernquist 
1992).

The feedback mechanism due to the energy released by the AGN, such that
studied for cooling flows in elliptical galaxies (Ciotti
et al 2001) might not be efficient here, because of the low luminosity of
LLAGN, and also the low coupling with the gas in a disk. Instead, other
intrinsic feedback mechanisms exist, related to the dynamical mechanisms
themselves that drive the gas to the center.
In these cases, the M$_{bh}-\sigma$ relation could be explained only
with the feedback mechanisms related to bars, that both can be responsible
for bulge and BH formation (e.g. review in Combes 2001).

The demographics of nearby AGN reveals that LLAGN exist in
about 40\% of all galaxies, and they
tend to lie in early-type galaxies (Terlevich et al 1987, 
Moles et al 1995).
In an optical spectroscopic survey of 486 nearby galaxies, 
Ho et al (1997) detected
420 emission-lines nuclei (86\% detection rate). Half of these objects can
be classified as HII or star-forming nuclei, and half 
as some kind of AGN: Seyfert,
LINERs and transition objects LINER/HII. A signature of Broad Line Region is
found in 20\% of the AGN, while Seyfert nuclei reside in about 10\% of
all galaxies. AGNs are found predominantly in luminous, early-type galaxies,
while HII nuclei are in less luminous late-type objects, which is 
compatible with the M$_{bh}-\sigma$ relation.

Bars are present in roughly two thirds of spiral galaxies. The frequency
of bars and non-axisymmetries has recently been quantified in details
from near-infrared surveys (Block et al 2002, Laurikainen et al 2004),
and the fraction of bars does not seem to vary with redshift (e.g. Jogee
et al 2004). Since bars are observed to have a suicidal behaviour in
spiral galaxies with gas (e.g. Hasan et al 1993, Friedli \& Benz 1995),
bars must be reformed to explain their frequency (Bournaud \& Combes 2002).
 The bar is destroyed by two main
mechanisms: first the central mass concentration built after the
gas inflow, destroys the orbital structure sustaining the bar,
scatters stellar particles and pushes them on chaotic orbits (Hozumi \& 
Hernquist 1999; Shen \& Sellwood 2004). Second, the gas inflow itself weakens
the bar, since the gas loses its angular momentum to the stars
forming the bar (Bournaud \& Combes 2005). This increases
the angular momentum of the bar wave, in decreasing
the eccentricity of the orbits.

\begin{figure}[h]
   \centering
   \includegraphics[width=9cm]{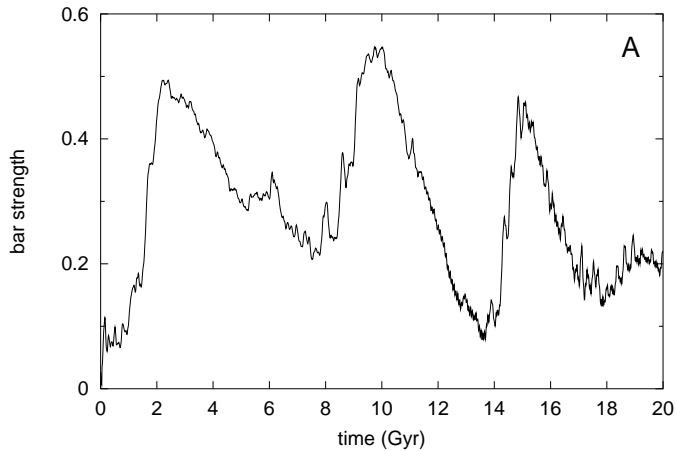}
      \caption{Evolution of bar strength in a simulated
spiral galaxy with gas accretion: several bar episodes
provide a flow of matter towards the center, to fuel
both the bulge and a central black hole (from Bournaud \& Combes 2002).}
       \label{fig2}
   \end{figure}

This bar destruction is reversible, and other bar episodes are driven
by external gas accretion, replenishing the gas disk
(cf Figure \ref{fig2}). 
A typical spiral galaxy is in continuous evolution, 
and must accrete gas all along its life, both to
maintain its star formation rate, and its spiral and bar structure.
The amount of gas required is able to double the galaxy mass in
about 10 Gyr. This gas cannot be provided by accretion of gas-rich
dwarf galaxies, since the interaction with companions 
would heat and destroy the disk. Instead, cold gas from
cosmic filaments must inflow to replenish the galaxy disk;
this can decrease temporarily the bulge-to-disk ratio,
making the disk more self-gravitating, and triggering
another bar instability. Several bar episodes
can succeed each other in a Hubble time, through this
dynamical feedback.
At each bar episode, both bulge and BH
grow in a similar manner, which explains the M$_{bh}-\sigma$ relation.

It is then more easy to understand the lack of correlation 
between the presence of bars and nuclear activity in spiral
galaxies.  The gas is driven to the very center only
intermittently, through the action of a secondary nuclear bar,
or even viscous torques, once the primary bar has
been dissolved by the main gas flow (Garcia-Burillo
et al 2005). The first gas flow is frequently stalled 
at the inner Lindblad resonance, responsible for a nuclear
starburst. Only when the bar has dissolved, can the gas 
fuel the nucleus. The activity of the nucleus can
occur in short episodes, which time scales are much shorter
than the bar formation and dissolution time scales,
which are of the order of 1 Gyr.

\begin{figure}[h]
   \centering
   \includegraphics[width=9cm]{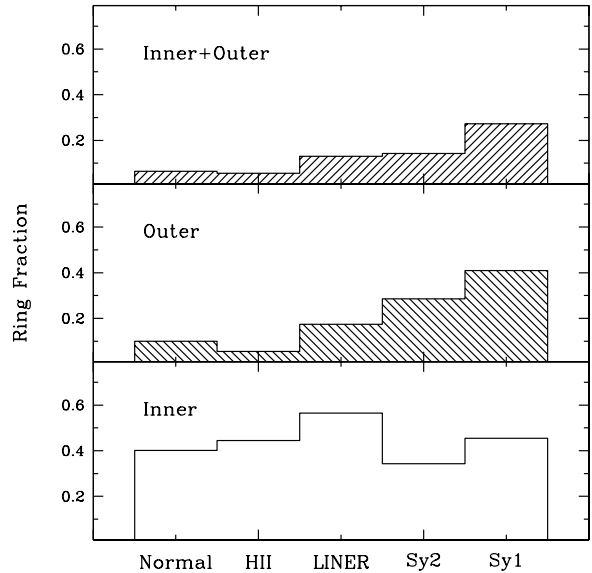}
      \caption{ Fraction of inner and outer rings in galaxies
as a function of their activity class. Rings are thought to
be formed at resonances with bars, near the corotation,
for the inner rings, and the OLR (Outer Lindblad Resonance)
for the outer rings.
Only objects with $v < 5000 km/s$ are shown (from Hunt \& Malkan 1999).}
       \label{hunt}
   \end{figure}

There are however some components, like resonant rings
in galaxy disks, which are the remnants of the presence
of bars in galaxies: once the bar has dissolved, the
rings survive for some more time, from stars formed
in the previously gaseous rings (e.g. Buta \& Combes 1996).
Unbarred galaxies observed with three resonant rings can be
considered as good evidence for the bar dissolution phenomenon.
 The presence of outer rings has been found to be predominant
in Seyfert galaxies (Hunt \& Malkan, 1999, and Figure \ref{hunt}).

\subsection{Other mechanisms}

Since the relation between BH and bulge mass
does not include the disk, ideas to involve
only a spheroidal system in the 
required angular momentum transfer, led to the radiation
drag from the bulge stars. The relativistic drag force
by the radiation from bulge stars is exerted
on the dusty gas ditributed spheroidally in the bulge.
 In practice, the radiation drag saturates
in case of optical thickness, but the optically
thin envelope of interstellar clouds is stripped,
loses angular momentum, and is accreted by the center
(Umemura 2001).
The M$_{bh}$/M$_{bulge}$ is then an universal constant
depending only on the energy conversion efficiency for
 nuclear fusion of hydrogen to helium.
 The efficiency of nuclear reactions in stars is
$\epsilon = 0.007$, which would predict a too important
ratio between BH and bulge mass.
The efficiency falls as 1/$\tau^2$, with $\tau$ the optical depth of the gas.
Also the radiation drag could be strongly subject to 
geometrical dilution.

The global scenario relates the  ultra-luminous infra-red
starburst phase and the formation of a quasar. During the first phase,
a large number of stars are formed, while the black hole is
still growing. In this phase, the BH-to-bulge mass ratio
is well below the present relation (such as has been proposed for
NLS1, Mathur 2000). Then at the end of the starburst phase,
the black hole has grown, and radiates at maximum
luminosity, in its QSO phase, while the optical thickness of
the interstellar medium decreases. Then the black hole 
ends up its growing phase, with the well-known
ratio between BH and bulge mass.
Today this mechanism is no longer inefficient, since elliptical galaxies
and bulges have no gas.
The mechanism has been quantified, with the radiative
transfer in a clumpy medium by Kawakatu \& Umemura (2002).
This idea has the advantage to explain why the BH mass 
is not proportional also to the disk mass, where radiation
drag loses its efficiency due to dilution and
optical thickness (Kawakatu \& Umemura 2004).

Also related to the formation of a starburst in a first phase,
is the formation of Super Star Clusters (SSC) in the
centers of galaxies (e.g. de Grijs et al 2001).
Sinking of these Super Star Clusters in their dark halo,
due to dynamical friction, has been proposed to
form cuspy stellar bulges (Fu et al. 2003); the merging of small BH associated
to clusters would provide a mass ratio
M$_{bh}$/M$_{bulge}$ = 10$^{-4}$ only, slightly
below what is observed.
Some intermediate mass black holes (IMBH) of masses
800-3000 M$_\odot$ would form easily in dense and young
star clusters (Portegies Zwart et al 2004b).

\section{Stellar cusps and cores and binary black-holes}

The supermassive black hole present in every spheroid, 
has a gravitational influence on its stellar environment. 
It can form a cusp, through gravitational attraction,
or a hole by swallowing the low angular momentum stars
in the neighbourhood, or flatten a core, through interaction
with another merging black hole. The observation of the stellar
profile in galaxy centers can then teach us the formation history
of the SMBH (e.g. Merritt 2004).

Observed is a well known dichotomy between massive and small ellipticals
(e.g. Lauer et al 1995):
\begin{itemize}

\item Cusps (steep power-law in stellar central density profile) are characteristic
of low-mass ellipticals, with disky isophotes and weak rotation;

\item Cores (flat central density profile) are found in high-mass galaxies, 
with boxy isophotes and no rotation.
\end{itemize}

\subsection{Formation of a cusp of stars around the black hole}

The density profile in the stellar component 
around a massive black hole depends on the relative value
of the 2-body relaxation time scale with respect to the Hubble time,
or more precisely with respect to the formation time of the black hole.

The relaxation time can be expressed by
$T_{rel} = \frac{V^3}{8 \pi G^2 m\rho(r) log(\Lambda)}$
where $V$ is the mean relative speed between the stars, $m$
the mean stellar mass, and $\rho(r)$ the volumic density in the
nucleus ($log(\Lambda)$ is the Coulomb parameter). It is well known that 
globally in a galaxy, the relaxation time is much longer than the 
Hubble time, it varies approximately as $0.2 (N/logN) t_c$, if $N$ is
the total number of stars in the system, and $t_c = R/V$ is the
crossing time.
However, the relaxation time becomes shorter than the Hubble
time in dense systems like globular clusters, and the nuclear 
stellar clusters may also approach these conditions (small $N$).
For the galactic center, with a volumic density of stars of
10$^7$M$_\odot$/pc$^3$, $<$V$^2>^{1/2}$  = 225 km/s,
this relaxation time is 3 10$^8$ yr.

Young (1980) has computed the adiabatic growth of a black hole
inside a nuclear stellar cluster: the growth rate is assumed to be
longer than the cluster dynamical time scale but shorter than the 
relaxation time scale. Then a stellar cusp forms, stars being
attracted by the black hole. The power-law profile has a slope 
larger then 2, up to 2.5. Two regimes can be distinguished,
according to the initial mass of the black hole, the slope
being larger for more massive black holes
(Cipollina \& Bertin, 1994).

\begin{figure}[h]
   \centering
   \includegraphics[width=9cm]{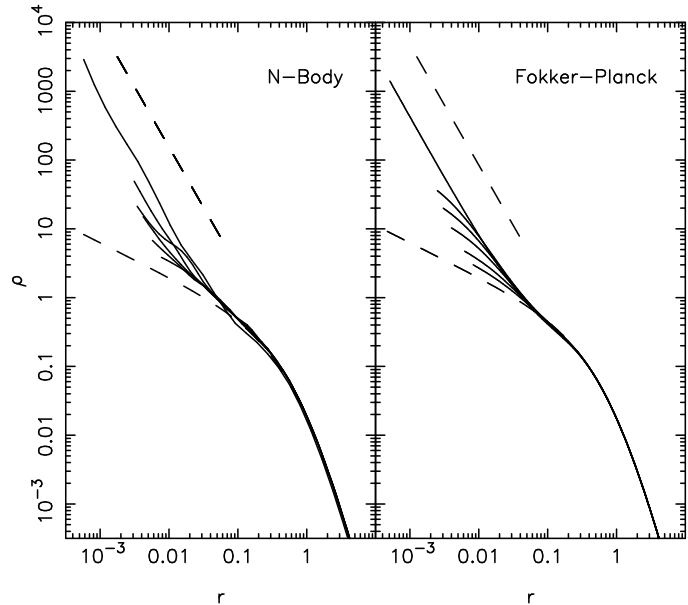}
      \caption{Evolution of the mass density profile around 
a massive black hole:
{\bf Left}: density from $N$-body  simulations, at
times $t/T_{rel}=0.07, 0.13, 0.2, 0.25, 0.33, 1.$ 
{\bf Right}: Densities predicted from the Fokker-Planck equation
at the same times.  The curves progressing from bottom to top,
are bracketted by the 
lower dashed curves at $t=0$ and the upper
dashed curves showing $\rho\propto r^{-7/4}$, the
asymptotic solution to the Fokker-Planck equation
(from Preto et al 2004).}
       \label{preto}
   \end{figure}

If 2-body relaxation can take place among the stars, then the
cusp is less pronounced, and the slope is 1.75 (7/4)
(Bahcall \& Wolf 1976).  N-body simulations 
can retrieve asymptotically this result,
cf Preto et al (2004) and Figure \ref{preto}.

The black hole can grow by swallowing the nearby stars, 
that have an angular momentum lower than
$J = (2 G M_{bh} r_t )^{1.2}$, with $r_t$ being the tidal
radius, beyond which a star is disrupted by the black hole.
After a dynamical time, if 2-body relaxation does not
refill these particles in the loss-cone, the black hole
will starve. In fact, the angular momentum can
diffuse faster than the energy (faster than the stellar relaxation time $T_{rel}$),
 and the low angular momentum stars are replenished faster,
which increases the accretion rate to
$T_{rel}$ (1-$e^2$), with $e$ the excentricity of the orbits
(Frank \& Rees 1976, Lightman \& Shapiro 1977).
It is possible that black holes less massive than
10$^7$ M$_\odot$ can be formed only through stellar accretion.

\subsection{Wandering of the black hole}

A massive compact object, like a black hole, embedded in a dense stellar system, 
experiences a multitude of gravitational encounters;
the total action on this body is composed of a slowly varying force, 
deriving from the smooth stellar system potential, and a rapidly 
fluctuating stochastic force due to discrete encounters with individual
stars. The motion of the black hole is then similar to that of a random 
walk (Chatterjee et al 2002, 2003), and this Brownian motion
has been invoked to counter the effect of the empty loss-cone,
and provide new stars to interact with. It is expected that
equipartition of energy is reached, so that the velocity acquired
by the black hole is small, even if the black hole interacts
with particles with high velocity dispersion. Numerical simulations
over-estimate this effect, since the number of particles
is far smaller than the realistic number.

The effect of wandering might be even more interesting on a
binary black hole (see next section). When interacting with a third
body, the binary can eject stars at large velocity. Also the dynamical
friction on a binary is less than on a single black hole.
Finally, when a black hole binary merges, the gravitational
waves emitted take away some momentum, producing a recoil of
the merged object. This hardly ejects it out of the galaxy (except
may be at high redshift, when the potential wells
are not deep enough), but can produce a large wandering
of the black hole, and a flattening of the cusp into a core
(Merritt et al 2004).

\subsection{Binary black holes}

\begin{figure}[h]
   \centering
   \includegraphics[width=9cm]{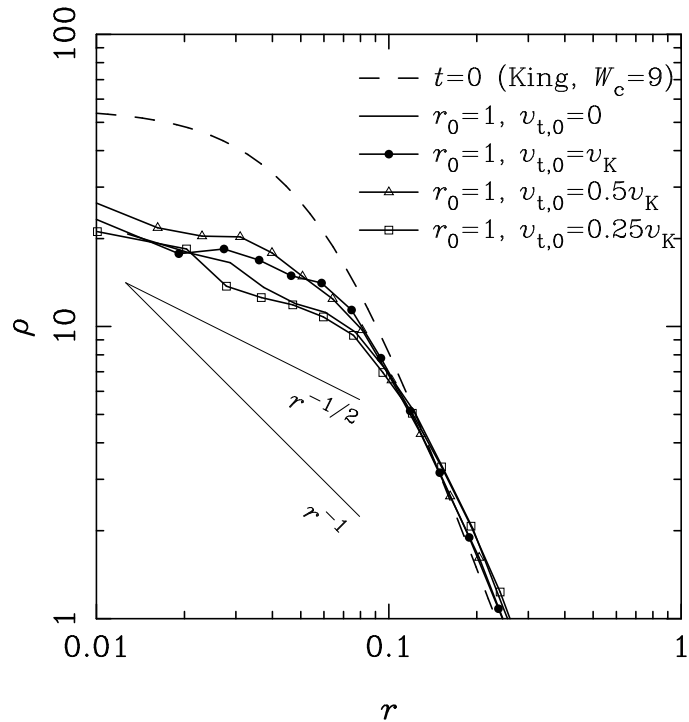}
\caption{Density profiles obtained through $N$-body simulations 
of the spiral-in of a massive BH, from Nakano \& Makino (1999).
The falling BH mass is 4\% of the galaxy mass.
The BH has initially only a tangential velocity
$v_{\rm t,0}$, function of the Kepler velocity $v_{\rm K}$, as
indicated on the figure.}
\label{nakano}
   \end{figure}

The formation of binary black holes in the centers
of galaxies is a natural prediction of the hierarchical
scenario, given the presence of a massive black hole in nearly
every galaxy.
The successive physical processes able to brake
the two black holes in their relative orbit have been
considered by Begelman et al (1980).
Each black hole sinks first toward the merger remnant center
through dynamical friction onto stars. A binary is formed;
but the life-time of such a binary can be much larger
than a Hubble time, if there is not enough stars to
replenish the loss cone, where stars are able to interact
with the binary.
Once a loss cone is created, it is replenished
only through the 2-body relaxation between stars,
and this can be very long ($T_{rel}$).
If the binary life-time is too long, another merger with
another galaxy will bring a third black-hole. Since a three-body
system is unstable, one of the three black-holes will be ejected
by the gravitational slingshot effect
(Saslaw et al 1974).

Numerical simulations have brought more precision in
the determination of the life-time of the binary,
although numerical artifacts have given rise to debates.
Ebisuzaki et al (1991) claimed that the life-time of
the binary should be much shorter if its orbit is excentric,
since then the binary can interact with more stars and
release the loss cone problem. The first numerical
simulations tended to show that orbit
excentricity should grow quickly through
dynamical friction (Fukushige et al 1992).
Mikkola \& Valtonen (1992) and others found
that the excentricity in fact grows only very slowly
(Quinlan 1996).

Numerical simulations suffer from a restricted number
of bodies N, and consequently of a large random velocity of the
binary (that should decrease in N$^{-1/2}$).
The binary then wanders in or even out of the loss cone,
and the effect of the loss cone depletion does not occur
(Makino et al 1993).
Also the 2-body relaxation time is shorter than
in the real system, contributing to replenish the cone.

More recent simulations, with increased number
of particles, have indeed shown that the 
hardening of the binary depends of the relaxation
time-scale, proportional to the number of particles
(Makino \& Funato 2004), and therefore another mechanism
is required to merge the binary, such as gas accretion.

The ejection out of the core of stars interacting with the binary
weakens the stellar cusp, while the
binary hardens. In addition, a sinking black hole
during a merger, contributes also to form a core
(Nakano \& Makino 1999, Figure \ref{nakano}). Gas dissipation, and star formation
in the central concentration formed, can restore
the cusp. This might explain the existence of cores
in the center of giant ellipticals, having experienced
multiple mergers, while small-mass systems have still a cusp.
 The computation of the deficit of stars in the central profiles,
and the formation of cores, appear in
agreement with observations (Graham 2004, Volonteri et al 2003b).

\begin{figure}[h]
   \centering
   \includegraphics[width=9cm]{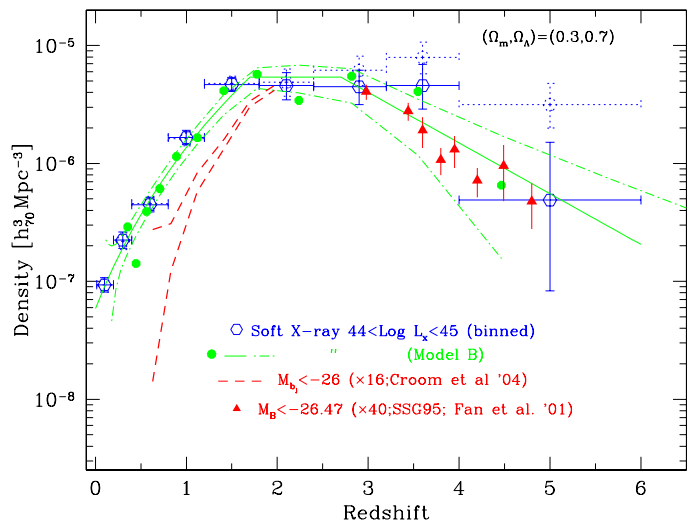}
      \caption{ Density of luminous QSO as a function of redshift,
from soft X-ray selected sources (empty circles), and optically selected ones 
(dash lines and triangles), from Hasinger (2004). The z $<$ 2 dash curve for optically selected 
QSO (M$_{bJ} < -26$) is from a combination of 2dF and 6dF surveys (Croom et al 2004). 
The triangles at  z $>$ 2.7 have been renormalised  from Schmidt et al (1995) and Fan et al 
(2001). }
       \label{Xbhgrow}
   \end{figure}

\section{History of accretion onto SMBH: X-ray constraints}

It is possible to relate the amount of energy produced in Active Galactic Nuclei, 
per comoving volume, to the mass density accreted by black holes, and therefore to 
their growth history, given an accretion efficiency $\epsilon$ 
 (e.g. Soltan 1982).
From the M$_{bh}$-M$_{bul}$ relation, the density of black hole
can be estimated to:
$$\rho_{bh} = 1.1 \,   10^6 (M_{bh}/M_{bul}/0.002) (\Omega_{bul}/0.002h^{-1}) M_\odot/ Mpc^3$$
as a function of the mass density $\Omega_{bul}$ in stellar spheroids.

The efficiency of conversion of mass into energy is generally adopted to be around 
$\epsilon$ = 0.1 for quasars, this value being an average between two extremes: for 
a Schwarzschild black hole, without rotation, this value is low ($\epsilon$  = 0.054), 
and for a Kerr black hole, with maximum rotation, it can reach $\epsilon$  = 0.37 
(Thorne 1974). It is possible to estimate the present growth of black holes by the 
optical QSO luminosity function, which yields an accreted mass density of
$$\rho_{accr} = 2 \, 10^5 \frac{0.1}{\epsilon}  M_\odot/ Mpc^3$$
 (Yu \& Tremaine 2002). This is however a lower limit, since most AGN light is absorbed 
at optical wavelengths.  The estimation from the far-infrared, assuming
a contribution of AGN to the FIR of 30\%, is:
$$\rho_{accr} = 7.5 \,  10^5 \frac{0.1}{\epsilon}  M_\odot/ Mpc^3$$
which confirms that the accretion radiation mainly does not get out
in optical light, but is re-radiated by dust.

It has been shown that the optically selected AGN correspond 
only to one third of the X-ray background (Barger et al 2003), which is now essentially 
resolved in individual sources, at least at energy lower than 2-5 kev (Worsley et al 2005). 
The accretion density estimated from the 
X-ray background has been estimated as high as 3-5 $10^5$ M$_\odot$/Mpc$^3$ (Salucci et al
 1999) and 6-9 $10^5$ M$_\odot$/Mpc$^3$ (Fabian \& Iwasawa 1999). These estimations have now been
updated to lower values. Taking into account the hard X-ray selected AGN, and 
their total corrected bolometric luminosity, Barger et al (2005) find a strong evolution 
with redshift of the AGN production rate, in (1+z)$^\alpha$, with $\alpha$ = 3.2 between 
z=0 and z=1. At higher redshifts the production decreases again (with $\alpha$ = -1), 
but the global integrated production is dominated by the z=1 objects. The deduced accretion 
density at z=0 is
$$\rho_{accr} = 4 \, 10^5 \frac{0.1}{\epsilon} M_\odot/ Mpc^3$$
 and about 40\% of this accretion 
density is due to the Broad-Line AGN, that are also the most powerful AGN (Steffen et al 2003).
The redshift evolution of the accretion rate is remarkably similar to the star formation 
history. Both histories reveal a downsizing effect, in the sense that the most active 
objects assemble mass in the early universe, and are no longer active now, while the 
remaining activity occurs now in the small-mass objects. Indeed, the most powerful and 
massive AGN observed at high redshifts have disappeared now, at the benefit of less powerful 
objects. This is also true for starbursts and ultra-luminous objects.
When compared with the density of black hole mass now in galaxies, estimated from the 
velocity dispersion of early-type galaxies determined from the Sloan Survey and the 
BH-mass to dispersion relation (Yu \& Tremaine 2002) or other estimations based on the 
mass density in the local universe and the M$_{bh}-\sigma$ relation
(Aller \& Richstone 2002, Marconi et al 2004), there is a good concordance 
with 
the mass expected from accretion luminosity, if the efficiency of accretion is 
$\epsilon$ =0.1. If the efficiency is higher, then there must exist obscured AGN, 
not counted in the above balance.

With deep X-ray surveys, it is now possible to draw quite precise conclusions on
the AGN redshift evolution, and luminosity functions (Hasinger 2004).
 There is a clear evolution of luminosity functions versus redshift,
which results from both number density evolution, and luminosity 
evolution. The evolution  with z depends strongly on luminosity.
For low-luminosity AGN,  the amplitude of number density evolution is less pronounced,
and the maximum occurs at low redshift, while bright quasars reveal a
factor 100 increase in density, and the peak occurs at higher z.
For L$_X$ = 10$^{42}$-10$^{43}$ erg/s, the peak is at z $\sim$ 0.5-0.7,
while it is at z$\sim$2 for  L$_X$ = 10$^{45}$-10$^{46}$ erg/s. 
This points towards a down sizing effect: rare and bright QSO
form very early in the universe, and then decline by two orders of
magnitude, while the more frequent low-luminosity AGN form later, and
decline by only a factor 10 in number. 

A clear decline at higher redshifts is now detected.
The comparison between the X-ray selected and optically-selected AGN
is illustrated in Figure \ref{Xbhgrow}. The comparison is
to be taken with caution, since the precise shape
of the number density evolution depends highly on 
luminosity.

\section{Conclusion}

The BH masses measured in local spheroids and spiral galaxies
with bulges, are tightly correlated to the central velocity dispersion,
and with more scatter to the bulge luminosity or mass. This has
been interpreted as a concomitant formation of stars in the bulge
and growth of the black hole at the center.
The BH growth occurs by a combination of external gas accretion, and 
coalescence of binary BHs, during galaxy mergers.
From the measured density of BH in the local universe, it is 
possible to deduce how many active galaxy nuclei have radiated
in the past, while accreting and growing these black holes.
The comparison of the optical, far-infra-red or X-rays outputs,
either in the form of point sources, or unresolved in the background,
with the local BH density constrains the radiating efficiency
of the AGN; the efficiency should be around
$\epsilon \sim 0.1$, in between that expected from Schwarzschild and Kerr black holes.

The discovery of high redshift (z$>$ 6) bright QSOs has posed
the problem of their formation in a short time-scale. This
problem appears related to the anti-hierarchical evolution
observed:  the brighter AGN form earlier, in a shorter time-scale,
while the low- luminosity AGN takes more time to form, and
reveals less evolution amplitude. Their peak number density
occurs at lower redshift than for bright QSO.
This behaviour might be explained by the much higher gas
density at high z, the higher merging rate and the shorter
dynamical time-scale.

The most massive black holes are confined in galaxy nuclei.
But there must exist an intermediate mass category for BH,
between the stellar-mass BHs and the supermassive ones.
Those IMBHs with masses around 10$^4$-10$^5$ M$_\odot$,
are not found particularly in nuclei, since they have not
been braked by dynamical friction. There appears now some evidence
of the existence of these IMBHs, but it is not yet clear
whether they might prolonge
the M$_{bh}-\sigma$ relation, at low masses.

Through the study of AGN demographics and several clues like
their cross-correlation length, the quasar lifetimes are determined to be
around a few 10$^7$ yrs; this is understood as the sum of all activity phases
in a single object, if there exist episodic recurrent activity. This life time
is luminosity dependent, being much longer for low luminosity AGN.
 
The relation between BH mass and bulge luminosity breaks down
for a certain category of AGN, the NLS1 which appear to have
formed their stars in the past at a higher rate than growing their
black hole. Evidence is found that nuclei in these objects are now accreting
mass much above the Eddington rate, although they are barely radiating around
the Eddington rate.  Gas accretion towards a starburst or a BH are then
not exactly concomitant, there could be time delays between the two
processes.

Can we trace back the formation of supermassive black holes
back to the early universe?  Small BH can form very early by
the collapse of Population III stars,  around z $\sim$ 20,
with masses of a few hundred solar masses. But the low efficiency
of dynamical friction on these small masses, make the BH growth
through merging unlikely. Most of the BH growth must then
be due to gas accretion.
 There must exist a large range in mass distribution of black holes
wandering all across galaxies; when accreting gas, these mini-quasars can
contribute to the reionization of the universe.

Different kinds of feedback processes have been invoked to 
account for the M$_{bh}-\sigma$ relation,  and in particular 
the energy released in AGN activity, QSO outflows, radiation,
that could self-regulate the gas accretion. Such processes are 
particularly conspicuous in the center of cooling flow clusters,
where gas re-heating regulate the cooling flow. Episodically,
the cooling gas fuels the central AGN, triggering a new
activity phase.
For low luminosity AGN, dynamical instabilities in galaxy disks,
like spirals and bars, are invoked to fuel the central nucleus,
and also self-regulate the gas accretion. Bars are destroyed
through gas inflow, and this could explain the apparent 
lack of correlation between nuclear activity and the presence
of  strong bars.

 The existence of binary black holes are a natural consequence 
of the hierarchical scenario of galaxy formation, if there exists
a supermassive black hole dormant in each nucleus.  The coalescence
of the binaries should occur relatively rapidly, to avoid the loss
of SMBH through 3-body interactions. Since dynamical friction
on bulge stars is not sufficient, the coalescence must be due to
gas accretion by the nucleus. During the hardening of the binary,
energy is given to the central stellar population, and any cuspy
density distribution can be flattened into a core, by this
dynamical heating. A cuspy stellar distribution can later reform
around the resulting single black hole.

In spite of large progress in massive black hole formation in the recent
years, many questions remain open, such as the evolution
of the M$_{bh}-\sigma$ relation with redshift, the local exceptions
to the relation (for instance galaxies like M33), the radiative efficiency
of the nucleus for a given accretion, etc..

\end{document}